\documentclass[11pt]{article}
\usepackage{bm}

\usepackage{epsfig}

\begin{document}

\newpage
\setcounter{page}{1}
\newpage
\section*{Torsion-balance probes of fundamental physics} 
\setcounter{page}{1}
\setcounter{section}{0}
{\bf E. G. Adelberger\newline
Department of Physics, University of Washington, Seattle WA 98195-4290}
\newline 
Modern torsion-balance experiments address a wide range of contemporary problems in fundamental physics. Because the experiments are are sensitive to extremely feeble forces they provide powerful constraints on many proposed extensions of the standard models of particle physics and gravity. Furthermore, the flexibility and relatively short time-scale of many of these experiments allow investigators to respond rapidly to new theoretical developments.
Here we outline the many motivations for this work, summarize some of the more interesting results and their implications, and project the future developments in this area. A previous review\cite{ad:09} covers much of this field in more detail.
\subsection*{Tests of the universality of free fall}
Conjectures about new scalar and vector fields permeate much modern thinking in particle physics and cosmology. Such quantum fields necessarily violate the universality of free fall (UFF)
because they couple to `charges' rather than mass, so that different electrically-neutral materials do not have the same free-fall acceleration. UFF tests 
are interesting because their extraordinary sensitivity allows one to see effects many orders of magnitude below the
gravitational scale that forms an irreducible background in conventional experiments. Such experiments therefore provide broad-gauge tests for new fundamental physics with length scales greater than 1 cm as well as testing the weak equivalence principle (WEP), a fundamental prediction of the standard model of gravity. 

It is conventional to parameterize UFF violation between electrically neutral atoms as a Yukawa interaction with range $\lambda$ that couples to generalized atomic `charges', 
$\tilde{q} = (Z\cos\tilde{\psi}+N\sin\tilde{\psi})$, where $Z$ and $N$ 
are the atom's proton and neutron numbers, and 
$\tilde{\psi}$
 specifies the details of the `charges'. 
This parameterization is exact for vector interactions, and a reasonable approximation for scalar fields. UFF tests compare the accelerations of two different materials in a composition dipole toward an attractor which can be a laboratory source, the earth, the  sun, our galaxy or the entire cosmos. Because there is always a value of $\tilde{\psi}$ for which the `charge' of any object vanishes, unbiased searches for new physics require UFF tests using at least 2 composition dipoles and 2 attractors. 
 The current state of this work (sensitive to forces $10^{13}$ times weaker than gravity) and some its implications are summarized in Ref.~\cite{wa:12}.
Differential accelerations toward the galactic center are particularly interesting because these laboratory experiments demonstrate that any non-gravitational, long-range interactions between hydrogen and galactic dark matter produces less than 10\% of the total acceleration\cite{wa:12}. Differential accelerations toward the sun, combined with the lunar laser-ranging EP test\cite{tu:07}, provide the best unambiguous test of the strong EP for gravitational self-energy.
Measurements of differential accelerations in the field of the earth yield high sensitivity to WEP-violating interactions with ranges between 1 m and infinity. An interesting application of differential accelerations of objects falling in the earth's field concerns speculations that antimatter may have different gravitational properties from normal matter. In field-theory language, this would imply that gravity has a vector component. But any vector component of gravity is so strongly constrained by WEP tests that the gravitational acceleration of antihydrogen is expected to differ from that of hydrogen by less than 1 part in $10^9$\cite{wa:12}.
\subsection*{Tests of the gravitational inverse-square below the dark-energy length scale}
It is useful to parameterize violation of the gravitational inverse-square law (ISL) in terms of a Yukawa interaction, but in this case one that couples to mass instead of a `charge'. The most sensitive current test\cite{ka:07} has shown that any ISL-violating interactions with gravitational strength must have a length scale less than about 50 $\mu$m. 
As detailed in a previous review\cite{ad:03}, particle and gravitational physics considerations provide compelling reasons
to search for violations of the gravitational inverse-square law (ISL) at the shortest possible length scales.  
The universe is apparently dominated by `dark energy' with a density $\rho_{\rm d} \approx 3.8$ keV/cm$^3$. 
This corresponds to a distance $\lambda_{\rm d}=\sqrt[4]{\hbar c/\rho_{\rm d}}\approx 85\mu$m that may represent a fundamental length scale of gravity\cite{dv:02} below which new phenomena may occur. These fall into 2 categories: new geometrical effects (extra-dimensions\cite{ar:98,dv:99}, etc) or extra forces from exchange of meV scale bosons\cite{ad:03}. These 2 categories can be distinguished by checking if the violation violates or obeys the UFF. An effect that obeys the UFF would constitute a `bombproof' signature of extra dimensions, an effect violating the UFF would be clear evidence for a new scale of particle physics (perhaps associated with M-theory's hundreds originally massless scalar particles with
`gravitational' scale couplings). Even if ISL violation is not seen, a rigorous upper-bound can be placed on the size of the largest extra dimension (currently 44 $\mu$m\cite{ka:07}. 
ISL tests also probe recent interesting speculations in non-gravitational particle physics. A particularly important example is the chameleon mechanism\cite{kh:04}.
If scalar bosons, for example, are given very small 
self-couplings then, in the presence of matter, essentially massless particles acquire effective masses that screen the interior of test bodies so that only a thin outer shell of the bodies is effective in sourcing or responding to the scalar field\cite{gu:04}. This essentially destroys the experimental limits on such bosons derived from astronomical and conventional laboratory  EP and ISL tests. However,  the test bodies in recent ISL tests are small enough to probe chameleons that couple to matter and to themselves with gravitational strength\cite{up:06}. Upadhye, Hu and Khoury\cite{up:12} recently noted that the 2007 ISL test \cite{ka:07,ad:07} excludes almost all ``chameleon field theories whose quantum corrections are well controlled and couple to matter with nearly gravitational strength regardless of the specific form of the chameleon potential''. They argue that a two-fold improvement in the minimum distance probed
would test {\em all} such theories. The next generation of torsion-balance experiments should reach this sensitivity.  
\subsection*{Planck-scale tests of Lorentz-symmetry violating and non-commutative geometry scenarios and searches for novel spin-dependent interactions}
Conventional EP and ISL experiments use unpolarized test bodies and attractors and are
completely insensitive to the purely spin-dependent forces such as those arising from the first-order exchange of unnatural parity ($0^-$, $1^+$, {\em etc}.) bosons. Experiments with electron-spin polarized pendulums and attractors probe such interactions and also provide a means to test for preferred-frame effects involving intrinsic spin. 
Dobrescu and Mocioiu\cite{do:06} have enumerated the kinds of potentials that can arise from one-boson exchange, constrained only by rotational and translational invariance. Most of these involve intrinsic spin. Perhaps the best motivated of these interactions is the spin dipole-dipole interaction that occurs in
theories with symmetries that are spontaneously broken at high energies\cite{ki:87} as well as in
torsion gravity, an extension of GR that arises in attempts to construct a gauge theory of gravity\cite{sh:02}. 
The masses and couplings of pseudo-Goldstone bosons created during spontaneous symmetry-breaking have the remarkable property that their masses as well as their coupling strengths are inversely proportional to the symmetry-breaking scale. Long-range (light exchange particle), ultra-feeble interactions are exactly the regime in which torsion pendulums excel.
Kosteleck\'y and collaborators\cite{co:97,ko:09} have developed a widely used theoretical framework, the Standard Model Extension (SME), for analyzing possible preferred-frame effects. Again, most of these involve spin.
Non-commutative geometry scenarios have received renewed interest from string theorists\cite{hi:02,an:01}.  This scenario predicts that a lepton spin will prefer to point in some direction in inertial space. 
The Eot-Wash group  
developed a torsion pendulum that contains  $\sim 10^{23}$ electron spins with essentially no external magnetic field\cite{he:08} as well as spin sources based on the same technology.
The spin pendulum was placed in a rotating torsion balance and used to searched for torques on the spins that tracked the orientation of the apparatus
relative to celestial coordinates or to an array of spin sources fixed in the lab. 
Sensitive searches for preferred-frame effects defined by the entire cosmos were made by checking whether the spins in the pendulum preferred to orient themselves in a direction fixed in inertial space, or if they had a generalized helicity defined by their velocity with respect to the rest-frame of the cosmic microwave background. Tight bounds were set on 9 combinations of Lorentz-symmetry violating violating SME parameters 
Finally, the effects of non-commutative space-time geometries were explored. In every case case, the constraints are interesting because of their extraordinary sensitivity. Upper bounds on some SME coefficients were between 4 and 5 orders of magnitude below the Planck-scale benchmark, while the bounds on non-commutative geometry are equivalent to an energy scale of $3\times 10^{13}$ GeV.
\subsection{Future prospects}
None of the above experiments have reached practical limits; all can be made more sensitive by challenging, but achievable, technical improvements. The UFF work is limited by thermal noise and changing gravity gradients. The thermal noise can be lowered
by using lower loss suspension fibers. Gravity-gradients can be continuously monitored, allowing corrections that greatly reduce this systematic effect. The physics reach can be extended by employing new test-body pairs that are `more different'. It is reasonable to foresee an order of magnitude improvement in the physics reach in the next decade. The short-distance ISL work is currently limited by both systematic effects and short-distance electrostatic noise.
Better metrology and surface preparation can extend the sensitivity to shorter length scales. However, the ISL violating signal diminishes rapidly as the length-scale falls so that it seems unlikely that current approaches can probe gravitational-strength interactions with length scales less than 10 $\mu$m or so. Substantial improvements in the some of the searches for new spin-dependent interactions are expected from new designs with higher symmetry.  

\end{document}